\documentclass[a4paper]{article}

\usepackage{graphicx}

\usepackage{cite}

\begin{document}

\begin{center}
{\Large \bf
Model Building of Metal Oxide Surfaces and 
} \\
{\Large \bf
Vibronic Coupling Density as a Reactivity Index: 
} \\
{\Large \bf
Regioselectivity of CO$_2$ Adsorption on Ag-loaded Ga$_2$O$_3$
}\\
~\\
{\large
Yasuro Kojima$^1$, 
Wataru Ota$^{1,2}$, 
Kentaro Teramura$^{1,3}$, 
Saburo Hosokawa$^{1,3}$,
Tsunehiro Tanaka$^{1,3}$, 
Tohru Sato$^{1,2,3}$*
}
~\\
{\large
\textit{
$^1$ Department of Molecular Engineering, Graduate School of Engineering, Kyoto University, Nishikyo-ku, Kyoto 615-8510, Japan
~\\
$^2$ Fukui Institute for Fundamental Chemistry, Kyoto University, Sakyo-ku, \\Kyoto 606-8103, Japan
~\\
$^3$ Unit of Elements Strategy Initiative for Catalysts \& Batteries, Kyoto University, Nishikyo-ku, Kyoto 615-8510, Japan
}
}
~\\
~\\
{\large
(Dated: Augast 23, 2018)
}

\renewcommand{\thefootnote}{\fnsymbol{footnote}}
\footnote[0]{* Corresponding author at: Fukui Institute for Fundamental Chemistry, Kyoto University, Takano Nishihiraki-cho 34-4 Sakyo-ku, Kyoto 606-8103, Japan. \\
\ \ \ \ \ \ \ \ Email address: tsato@scl.kyoto-u.ac.jp}
\end{center}

~\\

\begin{center}
	{\Large Abstract}
\end{center}
The step-by-step hydrogen-terminated 
(SSHT) model is proposed as a model for the surfaces of metal oxides.
Using this model, it is found that the vibronic coupling density (VCD) can be employed 
as a reactivity index for surface reactions.
As an example, 
the regioselectivity of CO$_2$ adsorption 
on the Ag-loaded Ga$_2$O$_3$ photocatalyst surface 
is investigated based on 
VCD analysis. 
The cluster model constructed by the SSHT approach
reasonably reflects the electronic structures of the Ga$_2$O$_3$ surface.
The geometry of CO$_2$ adsorbed on the Ag-loaded Ga$_2$O$_3$ cluster 
has a bent structure, 
which is favorable for its photocatalytic reduction to CO.

\newpage

\section{Introduction}

Heterogeneous catalysis, particularly for reactions 
between molecules and solid surfaces, has been extensively studied
\cite{Ertl1997}.
To design heterogeneous catalysts and understand of their mechanisms, 
the sites for molecular adsorption on the solid catalyst must be clarified.
The adsorption sites can be predicted theoretically
by finding the position of a molecule on a surface 
that has the lowest energy of all possible positions on the surface.
However, it is impractical to calculate all the energies 
because, in general, there are many adsorption sites 
for a molecule on a solid surface.
Therefore, a reactivity index 
to predict the adsorption sites 
based only on the information of solid surface is desirable.

Previously, we identified the regioselectivity 
of cycloaddition to fullerene 
\cite{Haruta2012_9702,Sato2012_257,Haruta2014_3510}, 
matallofullerene
\cite{Haruta2014_141}, 
and large polycyclic aromatic hydrocarbons
\cite{Haruta2015_590}
using the vibronic coupling density (VCD) as the reactivity index.
Vibronic coupling, the coupling between electron and nuclear vibrations, 
stabilizes a system by structural relaxation after charge transfer.
The VCD as a function of a position 
identifies the reactive sites as those
where the vibronic coupling is large.
It is expected that 
the VCD can be utilized as a reactivity index for systems
with various reactive sites, such as solid surfaces.

$\beta$-Ga$_2$O$_3$ is a heterogeneous catalyst 
that reduces CO$_2$ to CO using H$_2$ as a reductant
under photoirradiation
\cite{Teramura2008_191,Tsuneoka2010_8892}.
The selectivity of CO$_2$ reduction is increased 
by modifying $\beta$-Ga$_2$O$_3$ with Ag, which acts as a cocatalyst
\cite{Teramura2014_9906,Wang2015_11313,Wang2016_1025,Yamamoto2015_16810,Kawaguchi2018_459}.
H$_2$O, which is abundant, is used as the reductant in the Ag-loaded Ga$_2$O$_3$ system.
In this study, we applied VCD to the Ag-loaded Ga$_2$O$_3$ surface 
to show the effectiveness of VCD as a reactivity index 
for CO$_2$ adsorption on the surface.
This is the first report of the application of VCD to a solid surface 
and could provide the basis for 
extending the applicability of VCD to solid surfaces in general.

Reactivity indices, such as the frontier orbital density or VCD, strongly depend on the electronic structure of the frontier level.
When building a model for the surface reactions of metal oxides 
based on a bulk crystal structure, 
the treatment of dangling bonds strongly affects the electronic structure.
For instance, because hydrogen termination for dangling bonds involves electron doping, 
the frontier level is shifted by hydrogen termination.
In this study, to build a model for the subsequent calculations, 
we employed a step-by-step hydrogen-terminated (SSHT) approach 
to reproduce the experimental observations.

In Sec.~\ref{SEC2}, we describe the theory of vibronic coupling.
In Sec.~\ref{SEC3}, we describe the computational methods.
In Sec.~\ref{SEC4.1}, 
we present a method to build
a cluster model of the Ag-loaded Ga$_2$O$_3$ surface
by the SSHT approach.
In Sec.~\ref{SEC4.2}, 
we investigate the regioselectivity of CO$_2$ adsorption 
on Ag-loaded Ga$_2$O$_3$ cluster using VCD as the reactivity index.
Finally, in Sec.~\ref{SEC5}, we present the conclusions of this study.

\section{\label{SEC2} Theory}

In the early stage of chemical reactions, 
charge transfer occurs between reactants
by intermolecular orbital interactions.
Following charge transfer, 
the system is further stabilized by intramolecular deformation.
This structural relaxation is induced by vibronic coupling.
The vibronic coupling constant (VCC) $V$, 
which quantitatively evaluates the strength of vibronic couplings,
is defined by
\cite{Sato2008_758,Sato2009_99}
\begin{equation}
	V = \left( \frac{ \partial E_{{\rm CT}}}{\partial \xi} \right)_{{\bf R}_0},
\end{equation}
where $E_{{\rm CT}}$ is the total energy of the charge-transfer state, 
and ${\bf R}_0$ is the equilibrium geometry before charge transfer.
$\xi$ is the reaction coordinate along the nuclear vibration 
that gives the largest vibronic coupling
\begin{equation}
	\xi = \sum_{\alpha} \frac{V_{\alpha}}{\sqrt{\sum_{\alpha}|V_{\alpha}|^2}} Q_{\alpha},
\end{equation}
where $Q_{\alpha}$ is a normal coordinate and 
$V_{\alpha}$ is the VCC of vibrational mode $\alpha$.

The VCD $\eta$ is provided by the integrand of the VCC,
\begin{equation}
	V = \int d^3 {\bf r} \ \eta({\bf r}).
\end{equation}
Since $\eta({\bf r})$ is a function of the spatial coordinate ${\bf r}$, 
$\eta({\bf r})$ gives the local information about the VCC.
The VCD can be divided into electronic and vibrational contributions:
\begin{equation}
	\eta({\bf r}) = \Delta \rho({\bf r}) \times v({\bf r}).
\end{equation}
$\Delta \rho ({\bf r})$ is the electron density difference 
between neutral and charge-transfer states, 
and $v ({\bf r})$ is the potential derivative defined as 
the derivative of the potential acting on an electron from all the nuclei $u$ 
with respect to $\xi$.
The total differential of the chemical potential $\mu = \mu[N;u]$ 
which is a functional of the number of electrons $N$ and $u$ is given by
\cite{Sato2008_758}
\begin{equation}
	d\mu = 2 \zeta dN + \int \eta({\bf r}) d \xi d^3 {\bf r}, \label{dmu}
\end{equation}
where $\zeta$ is the absolute hardness.
In terms of the chemical reactivity theory proposed by R. G. Parr and W. Yang
\cite{Parr1984_4049,Parr1994}, 
the preferred direction for a reagent approaching a species 
is the one for which the initial $|d \mu|$ is the maximum.
The first term on the right-hand side of Eq.~(\ref{dmu}) 
is less direction sensitive than the second term.
Thus, the preferred direction can be said to be that for which 
the $\eta({\bf r})$ of a species is a maximum.

\section{\label{SEC3}Computational method}

For the calculations of the VCC and VCD, 
we first optimized the geometry of a neutral Ag-loaded Ga$_2$O$_3$ cluster 
and performed a vibrational analysis.
Then, we calculated the forces acting on the nuclei 
for the neutral optimized structure in a cationic state.
The charge-transfer state was chosen to be a cationic state 
because we assume that an electron is transferred 
from the Ag-loaded Ga$_2$O$_3$ surface to CO$_2$ 
in the reduction of CO$_2$.
Finally, we determined the adsorbed structure of CO$_2$
on the Ag-loaded Ga$_2$O$_3$ cluster by geometry optimization.
The computational level was set at the B3LYP/6-31G(d,p) level for the Ga, O, and H atoms 
and at the B3LYP/LANL2TZ level for the Ag atom.
The core electrons in the Ag atom 
were replaced by effective core potentials.
These calculations were performed using GAUSSIAN 09 
\cite{Frisch2013D,Frisch2013E}.
The VCC and VCD were calculated using our own code.

\section{Results}
\subsection{\label{SEC4.1}Cluster model of Ag-loaded Ga$_2$O$_3$ surface}

\begin{figure}[!ht]
\centering
    \includegraphics[width=0.7\hsize]{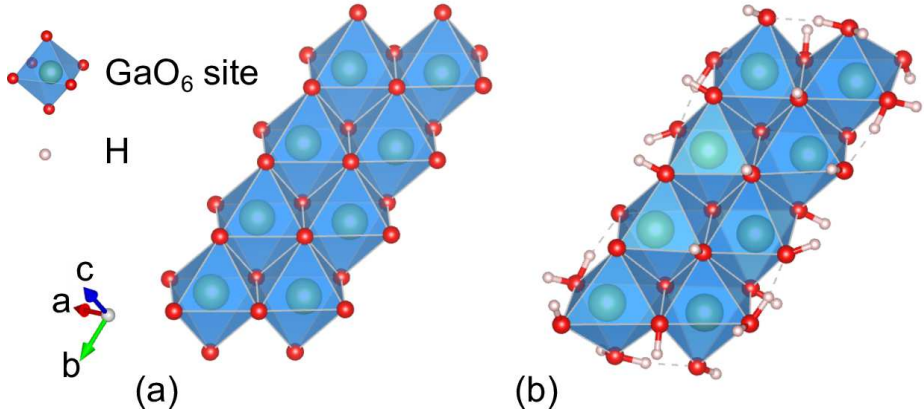}
    \caption{
	    Structures of the (a) bare and (b) SSHT model for the Ga$_2$O$_3$ surface
	    where Ga atoms are located at the octahedral sites formed by O atoms.
	     }
\label{FIG1}
\end{figure}

$\beta$-Ga$_2$O$_3$ consists of Ga atoms 
located at the octahedral and tetrahedral sites formed by O atoms
\cite{Geller1960_676,Aahman1996_1336}, 
as illustrated in Figure S1 in the Supplementary Material.
The octahedra share edges whereas the tetrahedra share corners in the $b$-axis direction.
The tetrahedra also share corners with the octahedra.
We expected that the electrons used for the reduction of CO$_2$ 
migrate through the octahedra shared edges.
In this study, eight adjacent octahedral sites 
are employed as a bare cluster model of the $\beta$-Ga$_2$O$_3$ surface 
(Figure~\ref{FIG1} (a)).

\begin{figure}[!ht]
\centering
    \includegraphics[width=0.6\hsize]{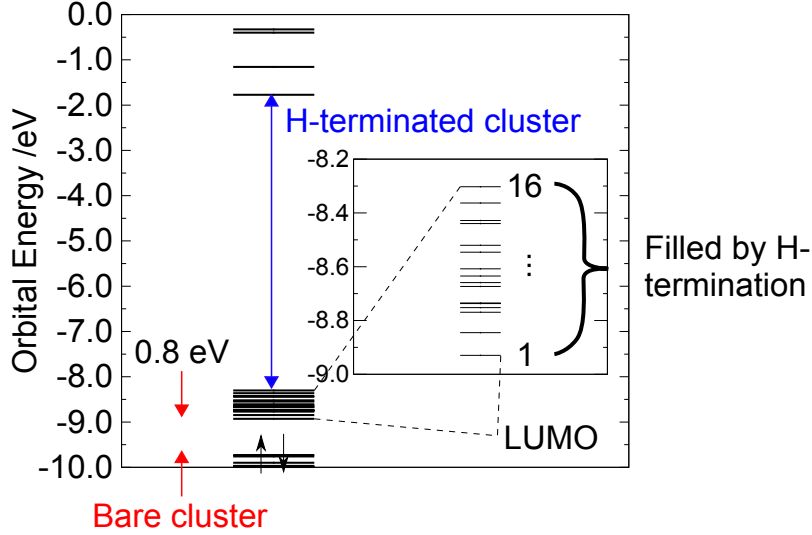}
    \caption{
	    Orbital levels of the bare cluster with an energy gap of 0.8 eV.
	    Dangling bonds at the O atoms 
	    are terminated with H atoms 
	    until the cluster model 
	    has an energy gap 
	    that agrees with the experimental value.
    }
	\label{FIG2}
\end{figure}

Figure~\ref{FIG2} 
shows the calculated orbital levels of the bare cluster.
The band gap of $\beta$-Ga$_2$O$_3$ has been
experimentally estimated to be 4.6 eV
\cite{Wang2016_1025}.
However, the energy gap of the bare cluster is 0.8 eV, 
which is much smaller than the experimental value.
This is because the occupied molecular orbitals 
become unoccupied when the cluster 
is cut from the crystal structure.
The bare cluster has reactive dangling bonds 
arising from the cleavage of O atoms.
The dangling bonds at the O atoms are terminated by H atoms
because it has been experimentally observed
that H atoms are adsorbed on the Ga$_2$O$_3$ surface
\cite{Tsuneoka2010_8892}.
The hydrogen termination, which involves electron doping,
shifts the frontier level.
The 16 unoccupied molecular orbitals must be occupied for the model 
to have a reasonably wide energy gap.
Thus, the 32 H atoms, i.e., 2 H atoms for each unoccupied molecular orbital, 
are step-by-step bonded to O atoms with large molecular orbital coefficients.
As a result, we obtained the hydrogen-terminated cluster 
with an energy gap of 5.4 eV after geometry optimization.
Figure S2 in the Supplementary Material
shows in detail the process of step-by-step hydrogen termination.
Hereafter, we refer to this hydrogen terminated cluster as the SSHT model 
for the Ga$_2$O$_3$ surface.
Figure~\ref{FIG1} (b) 
shows the optimized structure of the SSHT model.

\begin{figure}[!ht]
\centering
    \includegraphics[width=0.475\hsize]{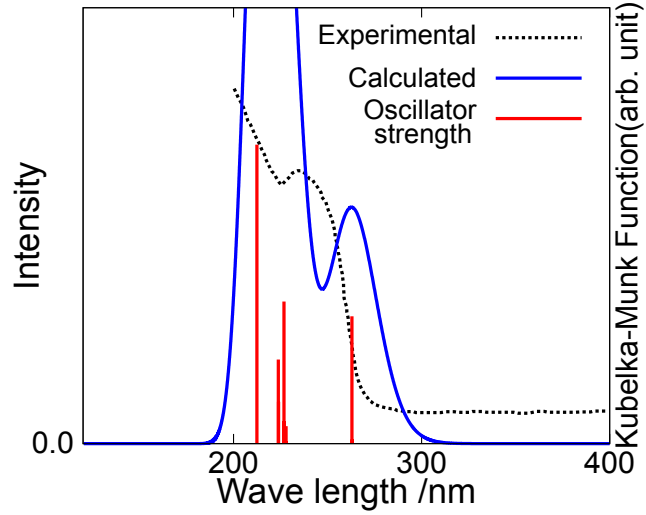}
  \caption{
	  Experimental (black dotted line)~\cite{Wang2016_1025} and 
	  calculated (blue line) diffuse reflectance spectra.
	  The red lines represent calculated oscillator strengths.
	   }
  \label{FIG3}
\end{figure}

The diffuse reflectance spectrum of the SSHT cluster model
is evaluated to examine its reliability.
The spectrum $g(x)$, which depends on the absorption energy $x$, 
is calculated from the oscillator strengths multiplied 
by the Gaussian distribution function,
\begin{equation}
	g(x) =  \sum_{i=1}^{10} \frac{f_i}{\sqrt{2 \pi \sigma^2}} 
	{\rm exp} \left( - \frac{ (x-u_i)^2 }{ 2 \sigma^2 } \right),
\end{equation}
where $f_i$ is the oscillator strength 
of a transition from S$_0$ to the Franck--Condon S$_i$ states, 
and $u_i$ is the excitation energy from S$_0$ to the Franck--Condon S$_i$ states. 
The value of $i$ is restricted between 1 and 10.
The values of $f_i$ and $u_i$ were calculated 
using time-dependent density functional theory.
Here, $\sigma^2$ is the variance of the Gaussian distribution function 
and was set to 0.05 eV$^2$.
Figure~\ref{FIG3} shows a comparison of 
the experimental and calculated diffuse reflectance spectrum 
\cite{Wang2016_1025}.
Although the calculated spectrum is shifted to the long-wavelength region 
with respect to the experimental spectrum, 
the two peaks of the calculated spectrum at 219 and 263 nm 
are also observed experimentally. 
Thus, the SSHT cluster is suitable for use
as a model for the Ga$_2$O$_3$ surface.

\begin{figure}[!ht]
\centering
    \includegraphics[width=0.4\hsize]{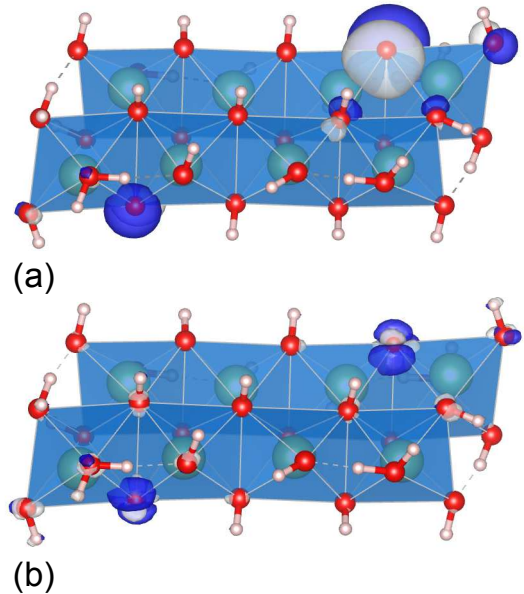}
  \caption{
	  (a) HOMO and (b) VCD $\eta({\bf r})$ of 
	  the SSHT cluster model for the Ga$_2$O$_3$ surface.
	  The HOMO and VCD are localized on the O atoms where H atoms are not bonded.
	  The isosurface values of HOMO and VCD are 
	  $3.0\times10^{-2}$ and $2.5\times10^{-5}$ a.u., respectively.
	  }
  \label{FIG4}
\end{figure}

It should be noted that, in the SSHT cluster model, 
H atoms are not bonded to all O atoms,
although all the dangling bonds are terminated by H atoms.
There are 20, 6, and 2 O atoms 
with which 1, 2, and 0 H atoms are bonded, respectively.
The O atoms without hydrogen termination have a pair of electrons, 
and, thus, act as Lewis bases.
This is supported by the highest occupied molecular orbital (HOMO) 
of the SSHT cluster model, as shown in Figure~\ref{FIG4} (a), 
which is strongly localized on the O atoms without hydrogen termination.
This result indicates that these O atoms donate electrons to the reactants.
The HOMO is doubly degenerate 
because the Lewis basic O atoms are 
located at both the front and back surfaces of the SSHT cluster model.
Figure~\ref{FIG4} (b) shows the VCD of the SSHT cluster model, 
which is also localized on the O atoms without hydrogen termination.
The stabilization arising from the structural relaxation after charge transfer 
is large at the sites where the VCD is localized.
Therefore, the Ag-loaded Ga$_2$O$_3$ surface 
is modeled 
by placing a single Ag atom on one of the Lewis basic O atoms 
in the SSHT cluster model.
The Cartesian coordinates of the SSHT cluster model for 
the Ga$_2$O$_3$ and Ag-loaded Ga$_2$O$_3$ surfaces
are given in Tables S1 and S2 of the Supplementary Material.

\subsection{\label{SEC4.2}Regioselectivity of CO$_2$ adsorption on Ag-loaded Ga$_2$O$_3$ cluster}

\begin{figure}[!ht]
\centering
    \includegraphics[width=0.6\hsize]{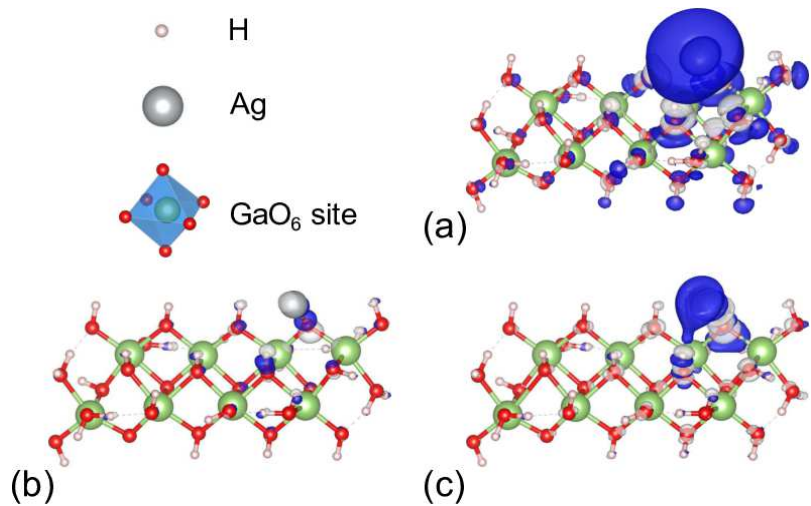}
  \caption{
	  (a) The electron density difference $\Delta \rho({\bf r})$, 
	  (b) potential derivative $v({\bf r})$, and 
	  (c) VCD $\eta({\bf r})$ of the Ag-loaded Ga$_2$O$_3$ cluster.
	  The isosurface values of 
	  $\Delta \rho ({\bf r})$, $v({\bf r})$, and $\eta ({\bf r})$ are
	  $10^{-3}$, $10^{-2}$, and $10^{-5}$ a.u., respectively.
	   }
	\label{FIG5}
\end{figure}

Figure~\ref{FIG5} shows the electron density difference $\Delta \rho({\bf r})$, 
the potential derivative $v({\bf r})$, 
and the VCD $\eta({\bf r})$ of the Ag-loaded Ga$_2$O$_3$ cluster.
Here, $\Delta \rho({\bf r})$, 
the electron density difference between the neutral and cationic states, 
is delocalized around the Ag atom 
because the electron is mainly extracted from the Ag atom.
$v({\bf r})$ is large at the Ag and adjacent O atoms.
Consequently, 
$\eta({\bf r})$, which is  
given by the product of $\Delta \rho({\bf r})$ and $v({\bf r})$,
is localized on the Ag atom as well as on the O atoms located near the Ag atom.
Since $\eta({\bf r})$ is distributed over 
the Ag atom and the O atoms at the surface of the Ga$_2$O$_3$ cluster, 
structural relaxation occurs between the Ag atom and the Ga$_2$O$_3$ cluster
following charge transfer.
This result implies that catalytic activity depends on the type of solid surface 
on which the Ag atom is loaded.

\begin{figure}[!ht]
\centering
    \includegraphics[width=0.5\hsize]{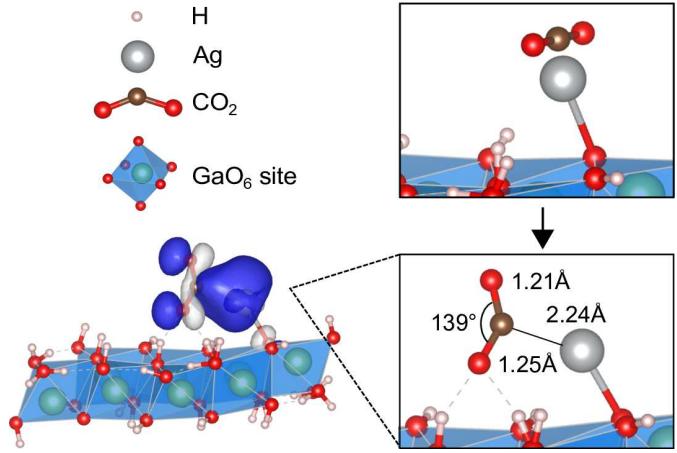}
  \caption{
	  Optimized structure and HOMO of CO$_2$ on the Ag-loaded Ga$_2$O$_3$ cluster.
	  The HOMO mainly consists of LUMO for bent CO$_2$ and Ag $s$ orbitals.
	  The isosurface value is 0.03 a.u.
	   }
  \label{FIG6}
\end{figure}

Geometry optimization 
is performed after the initial position of CO$_2$ is set above the Ag atom,
as shown in Figure~\ref{FIG6}.
The adsorbed structure of CO$_2$ 
is found to have an O-C-O angle of 139$^\circ$ and O-C distances of 1.21 and 1.25 \AA.
The adsorbed structure 
is obtained in the region where the VCD is localized.
The adsorption energy of CO$_2$ $E_{{\rm ad}}$ is defined by
\begin{equation}
	E_{{\rm ad}} = E_{{\rm CO_2}} + E_{{\rm cluster}} - E_{{\rm CO_2/cluster}},
\end{equation}
where $E_{{\rm CO_2}}$, $E_{{\rm cluster}}$, and $E_{{\rm CO_2/cluster}}$ 
are the energies of isolated CO$_2$, 
isolated Ag-loaded Ga$_2$O$_3$ cluster, 
and CO$_2$ on the cluster, respectively.
The $E_{{\rm ad}}$ of CO$_2$ 
is calculated to be 0.58 eV, 
indicating that the CO$_2$ with a bent structure is favorably adsorbed 
on the Ag-loaded Ga$_2$O$_3$ cluster.

\begin{figure}[!ht]
\centering
    \includegraphics[width=0.4\hsize]{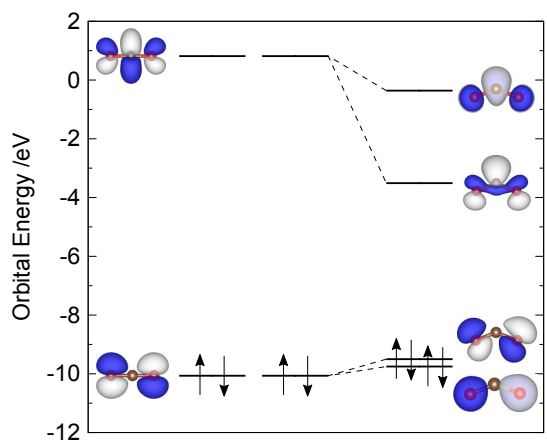}
  \caption{
	  Frontier orbital levels of CO$_2$ with a linear structure 
	  and that extracted from the structure in Figure~\ref{FIG6}.
	   }
  \label{FIG7}
\end{figure}

The reduction of CO$_2$ to CO is favored
when the LUMO level of CO$_2$ is low
\cite{Freund1996_225} 
because electrons are transferred 
from the Ag-loaded Ga$_2$O$_3$ surface to CO$_2$ for the reduction of CO$_2$.
Figure~\ref{FIG7} shows the frontier orbital levels of CO$_2$ 
with a linear structure and the structure extracted from Figure~\ref{FIG6}.
The LUMO level of bent CO$_2$ is lower than 
that of the linear structure although bent CO$_2$ is energetically unstable.
Thus, the Ag-loaded Ga$_2$O$_3$ surface 
on which bent CO$_2$ can be favorably adsorbed is suitable for CO$_2$ reduction.
As shown in Figure~\ref{FIG6}, 
the HOMO of CO$_2$ on the Ag-loaded Ga$_2$O$_3$ cluster 
is similar to the LUMO of bent CO$_2$ and Ag $s$ orbitals, 
leading to charge transfer 
from the Ag-loaded Ga$_2$O$_3$ cluster to CO$_2$.
The natural charges of the Ag atom 
are found to change from -0.03 to 0.48 upon CO$_2$ adsorption.

\begin{figure}[!ht]
\centering
    \includegraphics[width=0.7\hsize]{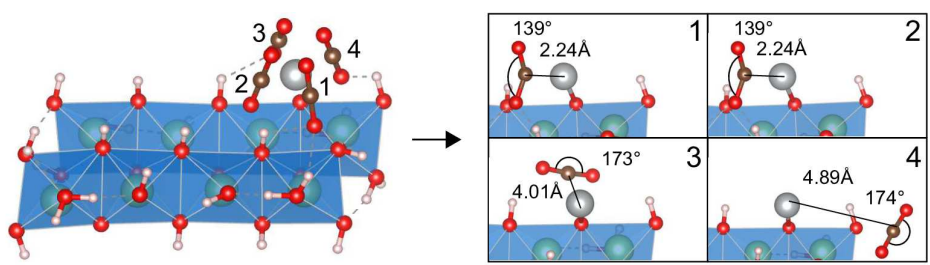}
  \caption{
	  Optimized structures obtained 
	  by changing the initial positions of CO$_2$ 
	  such that CO$_2$ surrounds the Ag atom.
	   }
  \label{FIG8}
\end{figure}

Geometry optimizations 
are performed by changing the initial positions of CO$_2$ 
to examine the dependence of the adsorbed structure.
The initial positions of CO$_2$ 
are prepared such that CO$_2$ surrounds the Ag atom 
where $\Delta \rho({\bf r})$ is distributed.
Although $\Delta \rho({\bf r})$ can also be used as a reactivity index, 
it tends to be delocalized compared to the VCD.
Figure~\ref{FIG8} shows the optimized structures 
obtained for each initial CO$_2$ position.
Optimized structures 1 and 2 in Figure~\ref{FIG8} 
are the same as that in Figure~\ref{FIG6}.
In structures 3 and 4, 
CO$_2$ has a linear structure that is unfavorable for the reduction of CO$_2$.
Furthermore, the $E_{{\rm ad}}$ of CO$_2$ for structures 3 and 4 
are calculated to be 0.47 and 0.30 eV, respectively.
These values are smaller than the values of $E_{{\rm ad}}$ for structures 1 and 2 of 0.58 eV.
Therefore, the adsorption of CO$_2$ 
in the region where the VCD is localized is 
advantageous for the CO$_2$ reduction, as well as 
being the most stable of the optimized structures.
Consequently, the regioselectivity of CO$_2$ adsorption is 
clearly indicated by the VCD because of the considerations of the vibronic contribution 
to the stabilization of the system
in contrast to $\Delta \rho({\bf r})$, which only considers the electronic contribution.
As shown in Figure S3 of the Supplementary Material,
the optimized structures obtained by placing CO$_2$
on Ga atoms are energetically unstable compared to the structure
obtained by the VCD analysis.

\section{\label{SEC5}Conclusion}

A cluster model of the Ga$_2$O$_3$ photocatalyst surface 
is constructed by terminating dangling bonds with H atoms.
The H atoms 
are bonded to O atoms with large orbital coefficients 
for each unoccupied orbitals 
such that the cluster model has an energy gap 
in agreement with the experimental values.
The O atoms without hydrogen termination 
act as Lewis bases.
This process of building a cluster model 
for metal oxide surfaces is termed as the
step-by-step hydrogen-terminated (SSHT) approach.
The vibronic coupling density (VCD) of the 
Ag-loaded Ga$_2$O$_3$ cluster 
is evaluated to identify the adsorption sites for CO$_2$.
Thus, the VCD is an effective reactivity index 
for determining the regioselectivity of CO$_2$ adsorption
on the Ag-loaded Ga$_2$O$_3$ surface.
We also found that CO$_2$ with a bent structure,
which is advantageous for photocatalytic reduction,
is adsorbed on the Ag atom.

\section*{Acknowledgments}

This work was supported by the Elements Strategy Initiative for Catalysts and Batteries (ESICB).
The calculations were partly performed using 
the Supercomputer Laboratory of Kyoto University and 
Research Center for Computational Science, Okazaki, Japan.

\bibliographystyle{elsarticle-num} 
\bibliography{refs.bib}

\begin{thebibliography}{10}
\expandafter\ifx\csname url\endcsname\relax
  \def\url#1{\texttt{#1}}\fi
\expandafter\ifx\csname urlprefix\endcsname\relax\def\urlprefix{URL }\fi
\expandafter\ifx\csname href\endcsname\relax
  \def\href#1#2{#2} \def\path#1{#1}\fi

\bibitem{Ertl1997}
G.~Ertl, H.~Kn{\"o}zinger, F.~Sch{\"u}th, J.~Weitkamp, Handbook of
  Heterogeneous Catalysis, 2nd ed., Wiley-VCH, Weinheim, 2008.

\bibitem{Haruta2012_9702}
N.~Haruta, T.~Sato, K.~Tanaka, J. Org. Chem. 77 (2012) 9702.

\bibitem{Sato2012_257}
T.~Sato, N.~Iwahara, N.~Haruta, K.~Tanaka, Chem. Phys. Lett. 531 (2012) 257.

\bibitem{Haruta2014_3510}
N.~Haruta, T.~Sato, K.~Tanaka, Tetrahedron 70 (2014) 3510.

\bibitem{Haruta2014_141}
N.~Haruta, T.~Sato, K.~Tanaka, J. Org. Chem. 80 (2014) 141.

\bibitem{Haruta2015_590}
N.~Haruta, T.~Sato, K.~Tanaka, Tetrahedron Lett. 56 (2015) 590.

\bibitem{Teramura2008_191}
K.~Teramura, H.~Tsuneoka, T.~Shishido, T.~Tanaka, Chem. Phys. Lett. 467 (2008)
  191.

\bibitem{Tsuneoka2010_8892}
H.~Tsuneoka, K.~Teramura, T.~Shishido, T.~Tanaka, J. Phys. Chem. C 114 (2010)
  8892.

\bibitem{Teramura2014_9906}
K.~Teramura, Z.~Wang, S.~Hosokawa, Y.~Sakata, T.~Tanaka, Chem. Eur. J. 20
  (2014) 9906.

\bibitem{Wang2015_11313}
Z.~Wang, K.~Teramura, S.~Hosokawa, T.~Tanaka, J. Mater. Chem. A 3 (2015) 11313.

\bibitem{Wang2016_1025}
Z.~Wang, K.~Teramura, Z.~Huang, S.~Hosokawa, Y.~Sakata, T.~Tanaka, Catal. Sci.
  Technol. 6 (2016) 1025.

\bibitem{Yamamoto2015_16810}
M.~Yamamoto, T.~Yoshida, N.~Yamamoto, T.~Nomoto, Y.~Yamamoto, S.~Yagi,
  H.~Yoshida, J. Mater. Chem. A 3 (2015) 16810.

\bibitem{Kawaguchi2018_459}
Y.~Kawaguchi, M.~Akatsuka, M.~Yamamoto, K.~Yoshioka, A.~Ozawa, Y.~Kato,
  T.~Yoshida, J. Photochem. Photobiol. A 358 (2018) 459.

\bibitem{Sato2008_758}
T.~Sato, K.~Tokunaga, K.~Tanaka, J. Phys. Chem. A 112 (2008) 758.

\bibitem{Sato2009_99}
T.~Sato, K.~Tokunaga, N.~Iwahara, K.~Shizu, K.~Tanaka, in: H.~K{\"o}ppel, D.~R.
  Yarkony, H.~Barentzen (Eds.), Vibronic Coupling Constant and Vibronic
  Coupling Density in The Jahn-Teller Effect: Fundamentals and Implications for
  Physics and Chemistry, Springer-Verlag, Berlin and Hidelberg, 2009.

\bibitem{Parr1984_4049}
R.~G. Parr, W.~Yang, J. Am. Chem. Soc. 106 (1984) 4049.

\bibitem{Parr1994}
R.~G. Parr, W.~Yang, Density-Functional Theory of Atoms and Molecules, Oxford
  University Press, New York, 1994.

\bibitem{Frisch2013D}
M.~J. Frisch~et al., Gaussian 09, Revision D. 01, Gaussian, Inc., Wallingford
  CT, 2013.

\bibitem{Frisch2013E}
M.~J. Frisch~et al., Gaussian 09, Revision E. 01, Gaussian, Inc., Wallingford
  CT, 2013.

\bibitem{Geller1960_676}
S.~Geller, J. Chem. Phys. 33 (1960) 676.

\bibitem{Aahman1996_1336}
J.~{\AA}hman, G.~Svensson, J.~Albertsson, Acta Cryst. C 52 (1996) 1336.

\bibitem{Freund1996_225}
H.-J. Freund, M.~W. Roberts, Surf. Sci. Rep. 25 (1996) 225.

\end{thebibliography}


\clearpage
\begin{thebibliography}{9}
\bibitem{Geller1960_676SM}
	S. Geller, J. Chem. Phys. 33 (1960) 676.
\end{thebibliography}

\newpage

\renewcommand{\figurename}{Figure S}
\renewcommand{\tablename}{Table S}
\setcounter{figure}{0}

\begin{center}
{\Large \bf
Supplementary Material
}\\
{\Large \bf
Model Building of Metal Oxide Surfaces and 
} \\
{\Large \bf
Vibronic Coupling Density as a Reactivity Index: 
} \\
{\Large \bf
Regioselectivity of CO$_2$ Adsorption on Ag-loaded Ga$_2$O$_3$
}\\
~\\
{\large
Yasuro Kojima$^1$, 
Wataru Ota$^{1,2}$, 
Kentaro Teramura$^{1,3}$, 
Saburo Hosokawa$^{1,3}$,
Tsunehiro Tanaka$^{1,3}$, 
Tohru Sato$^{1,2,3}$*
}
~\\
{\large
\textit{
$^1$ Department of Molecular Engineering, Graduate School of Engineering, Kyoto University, Nishikyo-ku, Kyoto 615-8510, Japan
~\\
$^2$ Fukui Institute for Fundamental Chemistry, Kyoto University, Sakyo-ku, Kyoto 606-8103, Japan
~\\
$^3$ Unit of Elements Strategy Initiative for Catalysts \& Batteries, Kyoto University, Nishikyo-ku, Kyoto 615-8510, Japan
}
}
~\\
~\\
{\large
(Dated: Augast 23, 2018)
}

\renewcommand{\thefootnote}{\fnsymbol{footnote}}
\footnote[0]{* Corresponding author at: Fukui Institute for Fundamental Chemistry, Kyoto University, Takano Nishihiraki-cho 34-4 Sakyo-ku, Kyoto 606-8103, Japan. \\
\ \ \ \ \ \ \ \ Email address: tsato@scl.kyoto-u.ac.jp}
\end{center}

\newpage

\begin{figure}[!ht]
\centering
    \includegraphics[width=0.4\hsize]{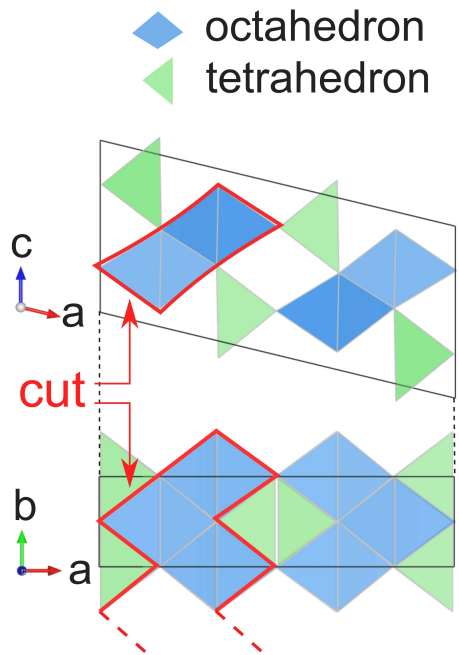}
  \caption{
	  $\beta$-Ga$_2$O$_3$ with lattice parameters 
	  $a$ = 12.23\AA, $b$ =3.04\AA, $c$ =5.80\AA, and 
	  $\beta$ = 103.7$^{\circ}$
	  \cite{Geller1960_676SM}.
	  Eight adjacent octahedral sites were 
	  used as the cluster model.
	   }
  \label{FIGS1}
\end{figure}

\begin{figure}[!ht]
\centering
    \includegraphics[width=0.9\hsize]{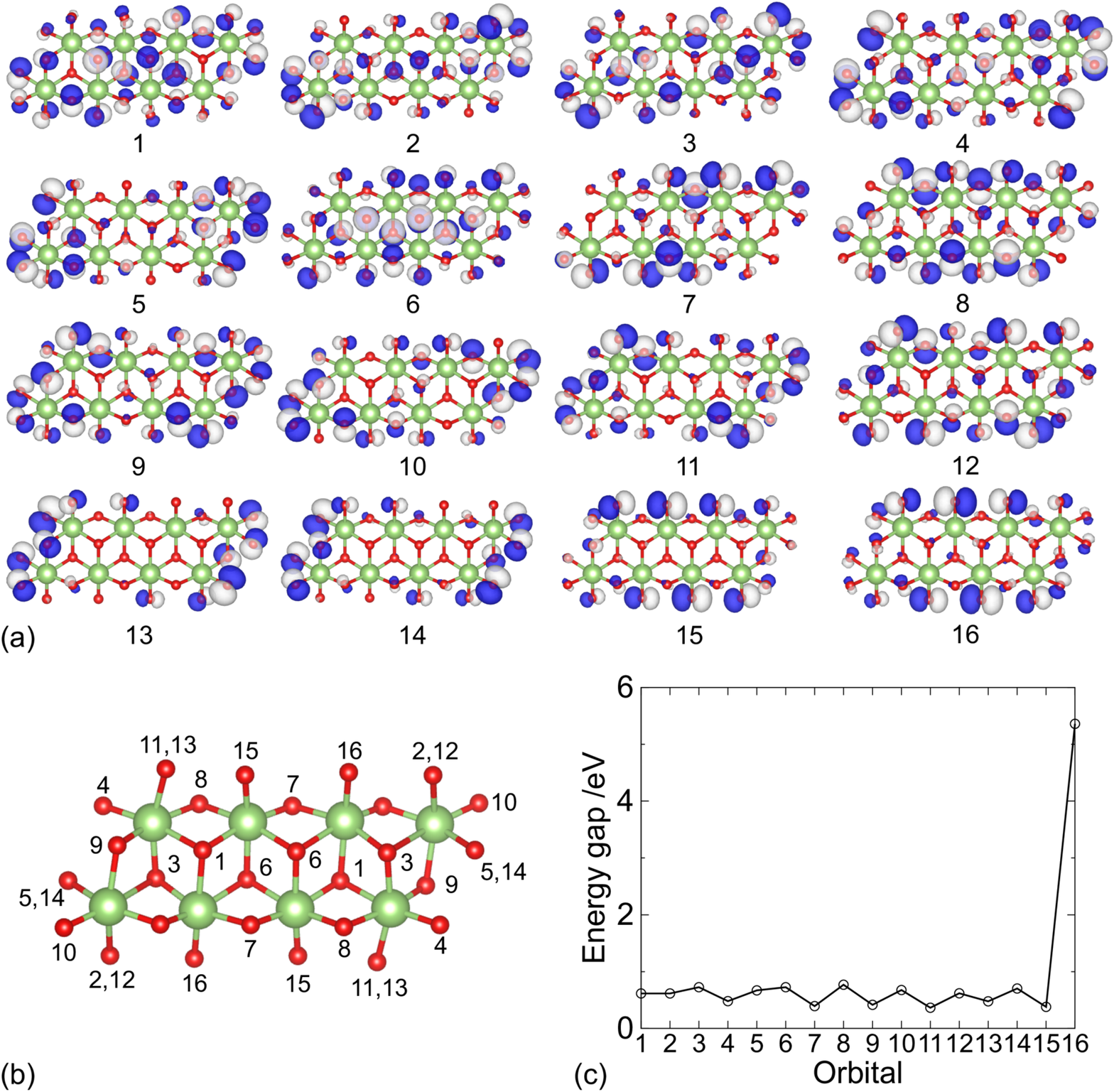}
  \caption{
	  (a) Orbital coefficients of the unoccupied molecular orbitals
	  of the bare cluster.
	  Numbers are given in ascending order of the orbital energy.
	  The orbital coefficients labeled 1 correspond to 
	  the LUMO of the bare cluster.
	  (b) The O atoms with large orbital coefficients 
	  for each unoccupied molecular orbitals.
	  Two H atoms were added step-by-step for each orbital.
	  (c) The energy gaps of the H-terminated clusters; 
	  these increase when orbital 16 is occupied.
	  The energy gaps were calculated by removing H atoms 
	  from the optimized SSHT model.
	  Because the bare cluster consists of 8 Ga and 28 O atoms, 
	  the sum of the oxidation numbers in the SSHT model 
	  is 0 after the introduction of 32 H atoms 
	  if we assume that the oxidation numbers of Ga, O, and H are 
	  +3, -2, and +1, respectively.
	   }
  \label{FIGS2}
\end{figure}

\begin{figure}[!ht]
\centering
    \includegraphics[width=0.9\hsize]{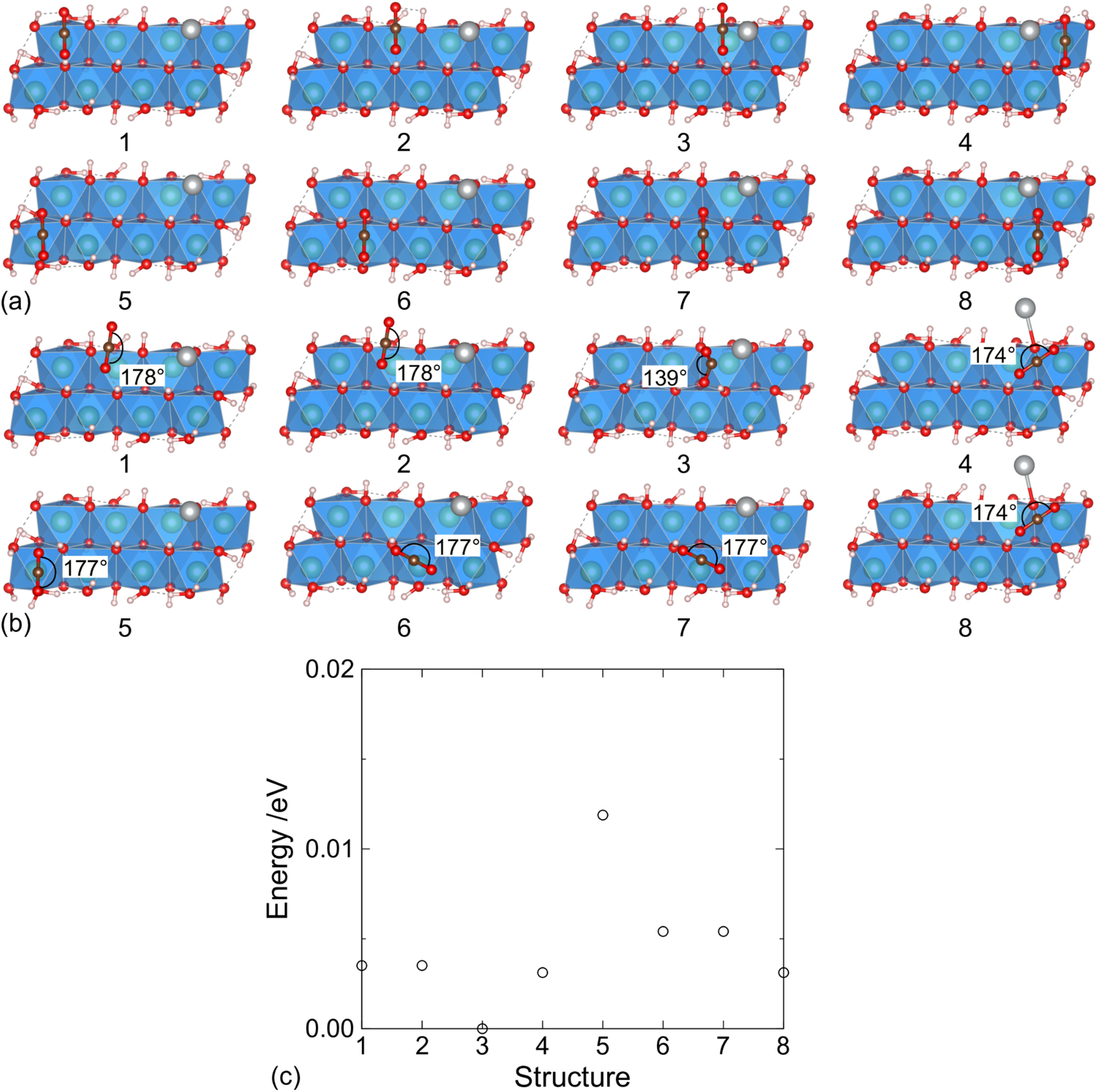}
  \caption{
	  (a) Initial configurations of CO$_2$ over the Ag-loaded Ga$_2$O$_3$ cluster 
	  where CO$_2$ is placed on the Ga atoms.
	  (b) Optimized structures and the values of the O-C-O angles. 
	  (c) Energies of the optimized structures subtracted from 
	  that of the structure predicted by VCD.
	  The calculated energy differences are zero for structure 3, 
	  which is the same as that predicted by VCD,
	  and positive for the other structures.
	   }
  \label{FIGS3}
\end{figure}

\begin{table}[!ht]
\caption{\label{table:Ga2O3} 
	Cartesian coordinates (\AA) of 
	the SSHT cluster model for the Ga$_2$O$_3$ surface.
}
\renewcommand{\arraystretch}{0.75}
\centering
\begin{tabular}{ccrrrccrrr} \hline
No. & Atom & 
\multicolumn{1}{c}{$x$} & \multicolumn{1}{c}{$y$} & \multicolumn{1}{c}{$z$} &
No. & Atom & 
\multicolumn{1}{c}{$x$} & \multicolumn{1}{c}{$y$} & \multicolumn{1}{c}{$z$} \\ \hline
1   & Ga &  -5.3237  & -0.8058  &  0.4860 &  
35  & O  &   6.8634  &  1.3493  & -1.3235\\
2   & Ga &  -2.4557  & -1.1803  &  0.0074 &
36  & O  &   5.2737  &  2.3049  &  1.0928\\
3   & Ga &   0.6034  & -1.4413  &  0.0528 &
37  & H  &  -2.3809  &  0.5814  &  1.7768\\
4   & Ga &   3.5952  & -1.7169  &  0.2520 &
38  & H  &   2.3806  & -0.5806  & -1.7762\\
5   & Ga &  -3.5954  &  1.7170  & -0.2521 &
39  & H  &  -5.7084  & -3.0520  & -0.6573\\
6   & Ga &  -0.6033  &  1.4414  & -0.0521 &
40  & H  &   5.7093  &  3.0523  &  0.6581\\
7   & Ga &   2.4559  &  1.1801  & -0.0061 &
41  & H  &  -3.5852  & -0.0498  & -2.1110\\
8   & Ga &   5.3241  &  0.8059  & -0.4877 &
42  & H  &   3.5810  &  0.0465  &  2.1123\\
9   & O  &  -4.9979  &  2.4166  & -1.3014 &
43  & H  &  -4.7369  &  2.8743  & -2.1082\\
10  & O  &  -1.9662  &  2.1832  & -1.2603 &
44  & H  &   4.7394  & -2.8780  &  2.1045\\
11  & O  &   0.9521  &  1.8286  & -1.1128 &
45  & H  &   6.4063  &  0.4370  &  1.7642\\
12  & O  &   3.8076  &  1.5368  & -1.1815 &
46  & H  &  -6.4097  & -0.4352  & -1.7654\\
13  & O  &  -3.8080  & -1.5367  &  1.1820 &
47  & H  &  -0.6458  & -0.3062  & -1.9339\\
14  & O  &  -0.9520  & -1.8284  &  1.1139 &
48  & H  &   0.6443  &  0.3063  &  1.9350\\
15  & O  &   1.9664  & -2.1840  &  1.2606 &
49  & H  &  -1.1551  & -2.7065  &  1.4561\\
16  & O  &   4.9990  & -2.4167  &  1.2993 &
50  & H  &   1.1557  &  2.7076  & -1.4526\\
17  & O  &  -5.2730  & -2.3053  & -1.0932 &
51  & H  &  -1.8420  &  3.1377  & -1.3470\\
18  & O  &  -2.6127  & -2.6522  & -1.2060 &
52  & H  &   1.8421  & -3.1388  &  1.3441\\
19  & O  &   0.6586  & -2.8876  & -1.1689 &
53  & H  &  -5.7097  &  1.6024  &  0.9589\\
20  & O  &   3.1106  & -3.3308  & -0.9146 &
54  & H  &   5.7086  & -1.6026  & -0.9594\\
21  & O  &  -4.9008  &  1.0750  &  1.0385 &
55  & H  &   6.6073  &  1.7510  & -2.1644\\
22  & O  &  -2.2019  &  0.6718  &  0.8309 &
56  & H  &  -6.6080  & -1.7503  &  2.1622\\
23  & O  &   0.7317  &  0.3336  &  0.9746 &
57  & H  &   3.6198  & -3.3427  & -1.7361\\
24  & O  &   3.5772  &  0.0145  &  1.1510 &
58  & H  &  -3.6202  &  3.3411  &  1.7368\\
25  & O  &  -3.5778  & -0.0151  & -1.1498 &
59  & H  &  -4.2915  & -2.5481  & -1.1593\\
26  & O  &  -0.7319  & -0.3337  & -0.9734 &
60  & H  &   4.2922  &  2.5476  &  1.1593\\
27  & O  &   2.2015  & -0.6715  & -0.8304 &
61  & H  &   2.0967  & -3.3029  & -1.1390\\
28  & O  &   4.9001  & -1.0747  & -1.0393 &
62  & H  &  -2.0973  &  3.3026  &  1.1393\\
29  & O  &  -6.8635  & -1.3497  &  1.3206 &
63  & H  &  -5.9264  &  1.0324  & -1.3045\\
30  & O  &  -6.3660  &  0.1727  & -1.0146 &
64  & H  &   5.9269  & -1.0315  &  1.3031\\
31  & O  &  -3.1113  &  3.3305  &  0.9150 &
65  & H  &  -0.1523  & -3.4128  & -1.1969\\
32  & O  &  -0.6595  &  2.8877  &  1.1698 &
66  & H  &   0.1507  &  3.4138  &  1.1979\\
33  & O  &   2.6135  &  2.6524  &  1.2064 &
67  & H  &  -2.3074  & -2.4901  & -2.1072\\
34  & O  &   6.3659  & -0.1715  &  1.0137 &
68  & H  &   2.3071  &  2.4913  &  2.1074\\ \hline
	\end{tabular}
\end{table}

\begin{table}[!ht]
\caption{\label{table:Ag-Ga2O3} 
	Cartesian coordinates (\AA) of 
	the SSHT cluster model for the Ag-loaded Ga$_2$O$_3$ surface.
}
\renewcommand{\arraystretch}{0.75}
\centering
\begin{tabular}{ccrrrccrrr} \hline
No. & Atom & 
\multicolumn{1}{c}{$x$} & \multicolumn{1}{c}{$y$} & \multicolumn{1}{c}{$z$} &
No. & Atom & 
\multicolumn{1}{c}{$x$} & \multicolumn{1}{c}{$y$} & \multicolumn{1}{c}{$z$} \\ \hline
1 & Ag  &  3.4721   &  0.9901   &  2.6839 & 
36& O   & -7.2861   & -1.4287   & -0.9215 \\
2 & Ga  &  5.0152   &  0.5699   & -0.4111 & 
37& O   & -5.5254   & -2.1278   &  1.4726 \\
3 & Ga  &  2.1368   &  1.0304   & -0.5291 &
38& H   &  2.0177   & -0.5219   &  1.4944 \\
4 & Ga  & -0.9058   &  1.3851   & -0.3358 &
39& H   & -2.8154   &  0.3464   & -1.9205 \\
5 & Ga  & -3.8739   &  1.7413   &  0.0414 &
40& H   &  5.4123   &  2.5685   & -1.8446 \\
6 & Ga  &  3.2169   & -1.9309   & -0.3957 &
41& H   & -6.0009   & -2.9146   &  1.1692 \\
7 & Ga  &  0.2420   & -1.5243   & -0.1789 &
42& H   &  3.3066   & -0.3939   & -2.4591 \\
8 & Ga  & -2.7873   & -1.1819   &  0.0537 &
43& H   & -3.6992   &  0.2321   &  2.0967 \\
9 & Ga  & -5.6848   & -0.8232   & -0.2682 &
44& H   &  4.1298   & -3.4437   & -2.1236 \\
10&  O  &  4.4839   & -2.8578   & -1.4458 &
45& H   & -4.8714   &  3.1355   &  1.8185 \\
11&  O  &  1.5526   & -2.4322   & -1.3223 &
46& H   & -6.5720   & -0.1702   &  2.0026 \\
12&  O  & -1.3805   & -1.9811   & -1.0918 &
47& H   &  6.1748   & -0.3780   & -2.5481 \\
13&  O  & -4.2244   & -1.6538   & -0.9690 &
48& H   &  0.2242   &  0.0160   & -2.2548 \\
14&  O  &  3.6392   &  1.5067   &  0.4331 &
49& H   & -0.8040   & -0.1199   &  1.7345 \\
15&  O  &  0.7329   &  1.8739   &  0.5495 &
50& H   &  0.9594   &  2.7516   &  0.8763 \\
16&  O  & -2.1660   &  2.2981   &  0.8613 &
51& H   & -1.6168   & -2.8966   & -1.2831 \\
17&  O  & -5.1896   &  2.5881   &  1.0922 &
52& H   &  1.3872   & -3.3842   & -1.3312 \\
18&  O  &  4.8647   &  1.8188   & -2.1216 &
53& H   & -2.0294   &  3.2545   &  0.8419 \\
19&  O  &  2.2856   &  2.2784   & -1.9727 &
54& H   &  5.4218   & -1.6329   &  0.7432 \\
20& O   & -1.0294   &  2.6581   & -1.7372 &
55& H   & -6.0690   &  1.5225   & -0.9928 \\
21& O   & -3.4568   &  3.1884   & -1.3476 &
56& H   & -7.1002   & -1.9333   & -1.7246 \\
22& O   &  4.6270   & -1.0912   &  0.6387 &
57& H   &  6.4974   &  1.8264   &  0.8840 \\
23& O   &  1.8832   & -0.6870   &  0.5391 &
58& H   & -4.0279   &  3.1172   & -2.1242 \\
24& O   & -0.9841   & -0.2594   &  0.7959 &
59& H   &  3.2559   & -3.1095   &  1.8587 \\
25& O   & -3.7963   &  0.1349   &  1.1444 &
60& H   &  3.8945   &  2.1306   & -2.1252 \\
26& O   &  3.2922   & -0.2994   & -1.5006 &
61& H   & -4.5463   & -2.3791   &  1.4924 \\
27& O   &  0.3703   &  0.1375   & -1.3085 &
62& H   & -2.4654   &  3.1023   & -1.6410 \\
28& O   & -2.5688   &  0.5479   & -1.0074 &
63& H   &  1.7408   & -3.2378   &  1.2512 \\
29& O   & -5.2753   &  0.9712   & -1.0647 &
64& H   &  5.4863   & -1.6314   & -1.7779 \\
30& O   &  6.6386   &  1.2583   &  0.1151 &
65& H   & -6.1247   &  1.2244   &  1.3303 \\
31& O   &  5.9832   & -0.7543   & -1.6802 &
66& H   & -0.2509   &  3.2215   & -1.8343 \\
32& O   &  2.7633   & -3.2885   &  1.0451 &
67& H   & -0.5095   & -3.2697   &  1.3902 \\
33& O   &  0.3263   & -2.8264   &  1.1916 &
68& H   &  1.8075   &  2.0646   & -2.7831 \\
34& O   & -2.8492   & -2.4997   &  1.4393 &
69& H   & -2.5080   & -2.2174   &  2.2972 \\
35& O   & -6.5896   &  0.3424   &  1.1826 \\ \hline
       \end{tabular}
\end{table}

\end{document}